# Observation of long-lived spin order in nanoconfined water

**Rohma Khan[1,2], Kang Xu[1], Nathaniel Jeffries[1], Ankit Bhardwaj[3,†], Boya Radha[3,4], Daniela Pagliero[1,*], and Carlos A. Meriles[1,2,*]**

[1]*Department of Physics, CUNY–The City College of New York, New York, NY 10031, USA.*
[2]*CUNY-Graduate Center, New York, NY 10016, USA.*
[3]*Department of Physics and Astronomy, The University of Manchester, Manchester, UK.*
[4]*Photon Science Institute and National Graphene Institute, University of Manchester, UK.*

[†]Present address: *Quantum Innovation Centre, Agency for Science, Technology and Research, Singapore 138634, Republic of Singapore.*

[*]E-mail: dpagliero@ccny.cuny.edu, cmeriles@ccny.cuny.edu

Liquids confined to nanometer-scale geometries exhibit behavior that departs markedly from their bulk counterparts, yet studying their dynamics under controlled conditions remains experimentally challenging. Here, we use nitrogen-vacancy (NV) center nuclear magnetic resonance (NMR) spectroscopy to probe water confined in 5.6 nm channels as a function of temperature. The system remains liquid throughout the investigated temperature range and exhibits strongly suppressed diffusivity, enabling direct detection of its $^{1}$H NMR spectrum. Occasionally, the proton resonance transforms into a doublet with a splitting of several tens of kilohertz, which we tentatively attribute to hyperfine interactions mediated by long-lived paramagnetic charge complexes, in turn seeded by solvated electrons optically injected during laser illumination. The intermittent appearance of this feature suggests a metastable state comprising a correlated population of charge-hydration complexes extending throughout the confined liquid.

Water confined within nanometer-scale environments is ubiquitous in biological[1], geological[2,3], and technological[4] systems. Under confinement, water can exhibit properties that differ markedly from those of the bulk liquid[5], including modified transport[6,7], phase behavior[8], and dielectric response[6,9]. Among the most intriguing manifestations of nanoconfinement are reports of strongly modified molecular dynamics, including substantial reductions in diffusivity and departures from bulk transport behavior[10-13]. Confinement has also been shown to alter the electrostatic environment experienced by water[14,15], as well as the structure of hydrogen-bond networks[5]. These effects are increasingly recognized to be interconnected: Local electric fields and charge accumulation can influence molecular mobility[11], while slow dynamics may in turn stabilize non-equilibrium electrostatic configurations[13,16].

Understanding this interplay remains a central challenge in the study of confined liquids, owing in part to the difficulty of simultaneously probing molecular transport and local electromagnetic environments on nanometer length scales. Recent advances in quantum sensing, particularly using nitrogen-vacancy (NV) centers in diamond, have begun to overcome these limitations[11,17-21]. Capable of detecting magnetic fields generated by nearby nuclear spins, NV centers provide direct access to the dynamics and local environment of liquids confined within nanometer-scale geometries[11]. This singular sensitivity has enabled studies of molecular transport, interfacial phenomena, and nanoscale fluid dynamics that are inaccessible to conventional NMR approaches.

Here, we employ NV-NMR spectroscopy to investigate water confined within ~5.6 nm channels over temperatures extending from ambient to near-freezing conditions. Throughout the explored range, the confined water remains liquid, and the proton NMR linewidth changes little, consistent with persistently suppressed molecular diffusion. In a subset of measurements, the proton resonance splits into a doublet separated by several tens of kilohertz. We first examine the constraints imposed by this observation on the microscopic state of the confined liquid and explore a tentative interpretation in terms of hyperfine interactions mediated by long-lived paramagnetic charge-hydration complexes.

To investigate nanoconfined water under variable-temperature conditions, we develop a temperature-controlled NV-NMR platform. Figure 1a shows the measurement geometry: Water is confined within ~5.6 nm channels formed by hBN layers deposited on a diamond substrate containing near-surface NVs. Optical spin initialization and readout are performed through the hBN structure — transparent throughout the excitation and detection windows — enabling the NVs to probe proton spins within the confined liquid. A Peltier cooler mounted beneath the sample holder enables operation from room temperature to near-freezing conditions. Integrated fluidic ports allow repeated filling of the channels via exposure to saturated water vapor[11] as well as control of the local water environment throughout the experiment (Supplementary

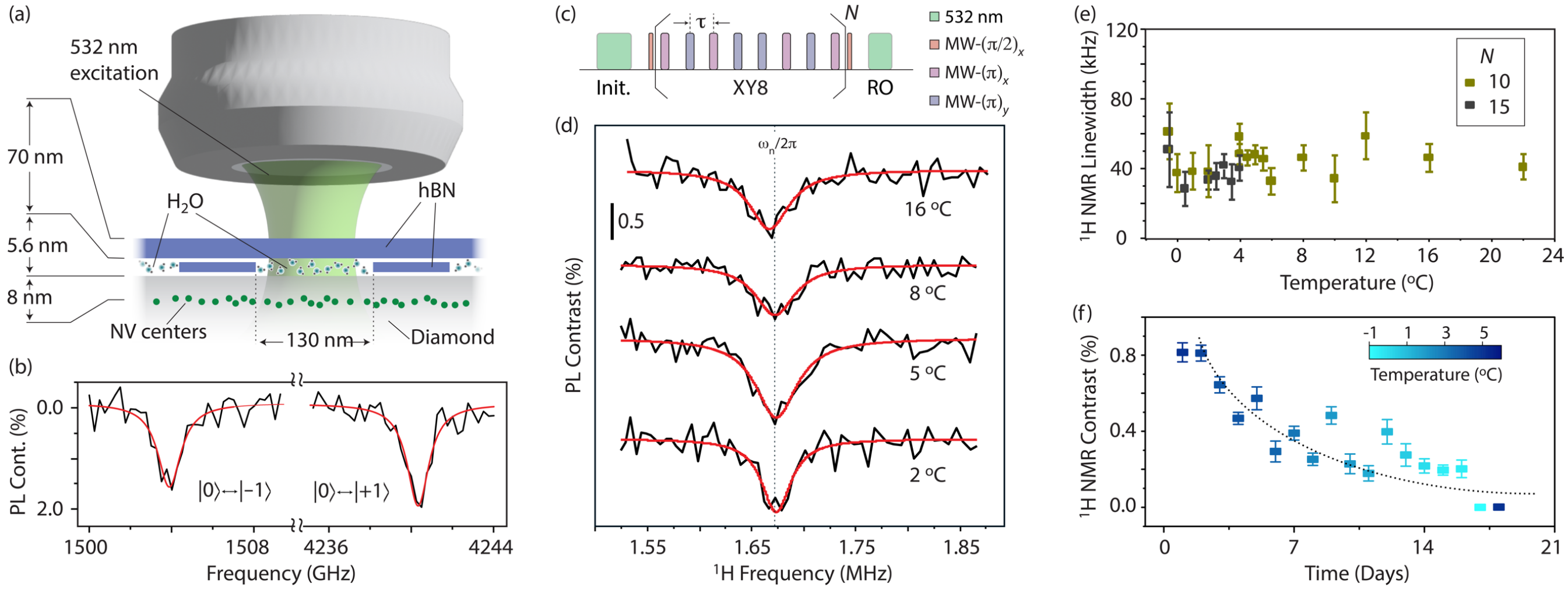

**Fig. 1: Temperature-controlled platform for nanoscale NV–NMR of confined water.** (a) Cross-sectional schematic of the device: near-surface NV centers in diamond probe $^1$H spins in water confined within 5.6-nm-thick hBN nanochannels beneath a 532-nm optical excitation spot. We control the position of a permanent magnet to create a magnetic field of variable amplitude and orientation and use a Peltier cooler to vary the sample temperature (not shown). (b) Representative NV optically detected magnetic resonance (ODMR) spectrum showing the $|0\rangle \leftrightarrow |\pm 1\rangle$ resonances at 4 ºC in the presence of a ~49 mT magnetic field aligned with one of the four possible NV axes. (c) Schematic of the dynamical decoupling protocol used for nanoscale NMR detection, based on an XY8 sequence with variable pulse number *N*. RO: Readout; Init.: Initialization. (d) Representative NV-detected, $^1$H NMR spectra at 39 mT and different temperatures. The red lines are Lorentzian fits to the experimental data (black). (e) Extracted NMR linewidth as a function of temperature for two representative 8-pulse-cycle numbers ($N = 10$ and $N = 15$). Within experimental uncertainty, the linewidth remains approximately constant across the investigated temperature range, including near and slightly below the nominal freezing point of bulk water. (f) Evolution of the NMR contrast over multiple days of low-temperature measurements in a single experimental run with the dashed line serving as a guide to the eye. The gradual decrease in signal amplitude is consistent with slow evaporation of water from the nanochannels, indicating that the confined water remains in a liquid-like state even at near-freezing conditions.

Material (SM), Section I).

Temperature-dependent measurements were performed under an applied magnetic field in the 35-50 mT range parallel to one of the NV crystallographic axes. Representative optically detected magnetic resonance (ODMR) spectra are shown in Fig. 1b for both the $|0\rangle \leftrightarrow |-1\rangle$ and $|0\rangle \leftrightarrow |+1\rangle$ transitions of the NV ground state manifold. Operation near the excited-state level anticrossing together with precise alignment of the magnetic field enables efficient polarization of the host $^{15}$N nuclear spin, resulting in ODMR linewidths of ~2 MHz, significantly narrower than typically observed in NV ensembles[22]. Averaging these resonances provides a direct measure of the zero-field splitting whose well-known temperature dependence allows for in-situ thermometry[23]. We attain good agreement between the temperatures extracted from this type of measurement and the applied Peltier control (SM, Section I).

We probe the confined liquid with the help of an XY8 dynamical decoupling sequence[24] (Fig. 1c). In this protocol, periodic microwave (mw) π-pulses render the NV centers selectively sensitive to magnetic fluctuations generated by nearby proton spins, with maximum response when the pulse spacing satisfies $2\tau = \nu_{\mathrm{H}}^{-1}$, where $\nu_{\mathrm{H}}$ is the proton Larmor frequency; we reconstruct the $^1$H NMR spectrum of the confined water simply by sweeping $\tau$.

Detecting proton NMR from liquid water with shallow NV centers is intrinsically challenging because molecular self-diffusion rapidly transports water molecules through the nanoscale sensing volume of individual NVs (~(8 nm)$^3$ in the present case), hence limiting the ability to encode in the NV spins a measurable phase change. For bulk water, the characteristic transit time is only a few nanoseconds, many orders of magnitude shorter than the microsecond interrogation times required for nanoscale NMR detection. Thus, the very observation of proton signals already implies molecular dynamics substantially slower than those of bulk water.

Representative spectra acquired at several temperatures are shown in Fig. 1d. Across the explored range, the proton resonance remains readily observable and exhibits a linewidth that varies only weakly with temperature. The extracted full-width-at-half-maximum values are summarized in Fig. 1e. Within the experimental scatter, the linewidth (~40 kHz) remains approximately constant over the investigated range, corresponding to effective molecular diffusivities on the order of $10^{-12}$ m$^2$ s$^{-1}$ or lower (see also SM, Section II), several orders of magnitude below that of ambient[11] or supercooled[25] bulk water (~$2\times10^{-9}$ m$^2$ s$^{-1}$ and ~$1.6\times10^{-10}$ m$^2$ s$^{-1}$, respectively). These observations indicate that the strongly suppressed liquid-state dynamics previously inferred at room

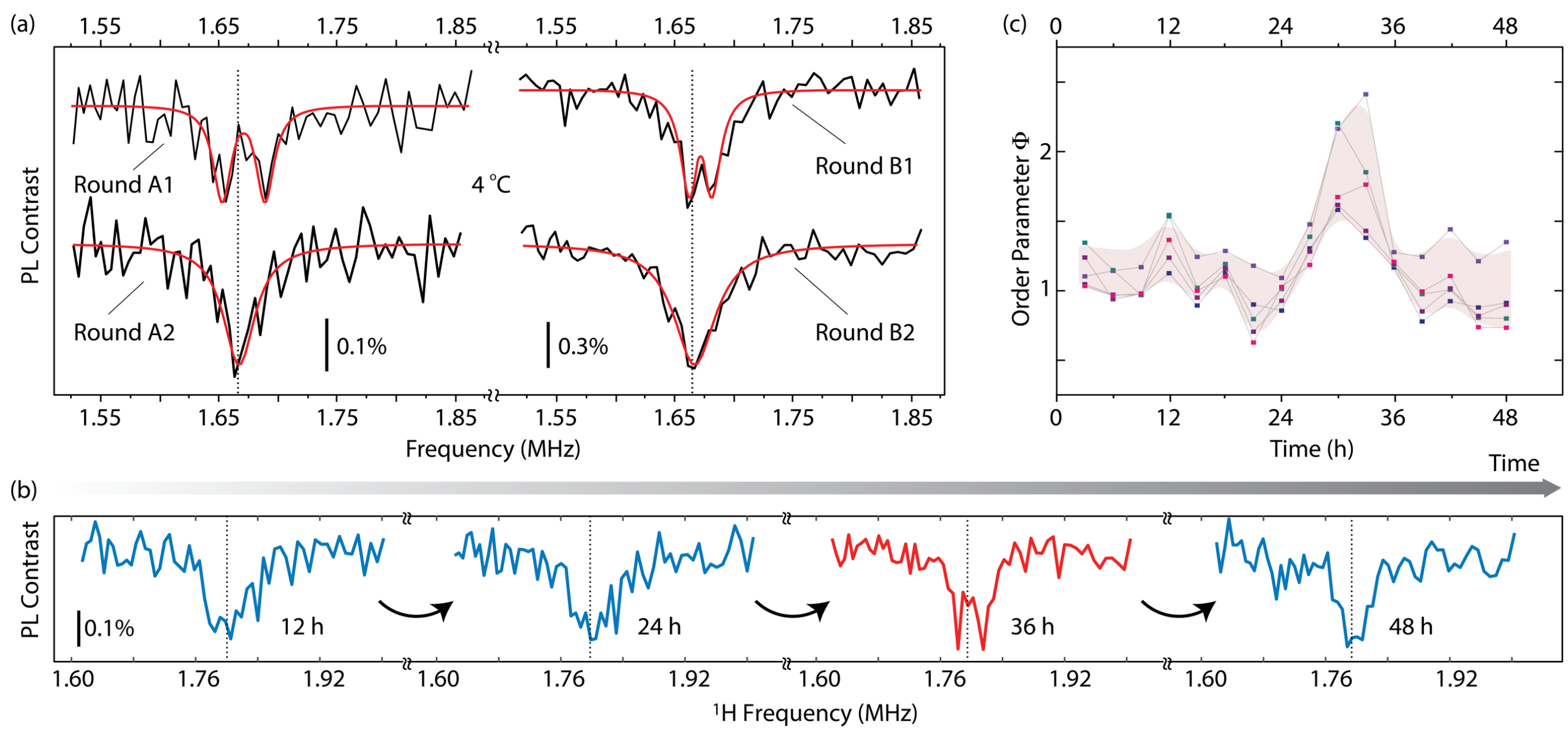

**Fig. 2: Intermittent doublet structure in the $^1$H NMR line near 4 °C.** (a) Representative NV-detected $^1$H NMR spectra at 4 °C as derived from the XY8-N protocol, showing examples of split (doublet-like) resonances. The left and right columns present results from two independent experimental runs (respectively, Rounds A and B), each obtained after resetting the microfluidic device by reintroducing water into the channels. In Round A, a pronounced doublet is observed in the first measurement (A1) but is not consistently recovered after subsequent temperature cycling (A2). In Round B, improved signal-to-noise ratio was achieved through enhanced measurement sensitivity shortly after a refill; however, the observed splitting is weaker than in Round A. Red curves are Lorentzian fits to the data (black). The vertical dashed lines indicate the anticipated position of the $^1$H NMR resonance. (b) Consecutive $^1$H NMR measurements acquired over 48 hours at 5 °C. The intermittent emergence and disappearance of the doublet suggest reversible switching between distinct microscopic configurations. (c) Time dependence of the order parameter Φ; colored dots indicate different fitting criteria, and the semitransparent band is a guide to eye (SM, Section II). The rise and fall of Φ mirrors the gradual appearance and disappearance of the transient split state shown in (b).

temperature[11] persist down to near-freezing conditions.

An important feature of the present nanochannel platform is that, unlike fully encapsulated confinement geometries, the channels can be repeatedly filled and emptied through the integrated microfluidic system. This capability makes it possible to directly monitor the evolution of the confined liquid through its proton NMR response. As shown in Fig. 1f, the signal amplitude after a refill gradually decreases over time, consistent with the progressive loss of water from the channels through evaporation. The proton signal can subsequently be restored by refilling the device, confirming that the observed decay originates from changes in channel occupancy. Importantly, the gradual reduction in signal amplitude is not accompanied by a change in linewidth, indicating that the molecular dynamics of the remaining confined water stay largely unchanged even as the channel filling fraction decreases (SM, Section II).

Although most measurements yield a single proton resonance, a subset of experimental runs — most frequently within the 4–5 °C window, but occasionally at temperatures as high as 14 °C as well, SM, Section II — reveals a qualitatively different spectral response. Representative examples are shown in Fig. 2a. In these instances, the $^1$H resonance appears as a resolved doublet with a splitting of several tens of kilohertz. Importantly, each pair of spectra was acquired within the same channel-filling cycle, with the split and unsplit lineshapes observed under virtually identical experimental conditions. The appearance of the doublet therefore reflects a spontaneous change in the magnetic environment experienced by the confined protons.

To investigate this behavior in greater detail, we monitored the proton resonance continuously over extended periods while maintaining the sample near 5 °C. As shown in Fig. 2b, the spectral lineshape evolves over timescales of many hours. Beginning from a single-line resonance, the spectrum develops a resolvable splitting before eventually returning to its original form. The abrupt nature of this evolution is best quantified in Fig. 2c, where we leverage consecutive measurements over the shortest possible time window and quantify the extent to which each spectrum exhibits doublet character using the order parameter Φ (SM, Section II). These observations best explain the intermittent appearance of the doublet in Fig. 2a, i.e., a transient configuration occupying only a limited portion of the measurement record. The associated timescale, extending over many hours, is orders of magnitude longer than characteristic molecular, spin, or hydrogen-bond relaxation times, thus suggesting the formation of a long-lived metastable configuration.

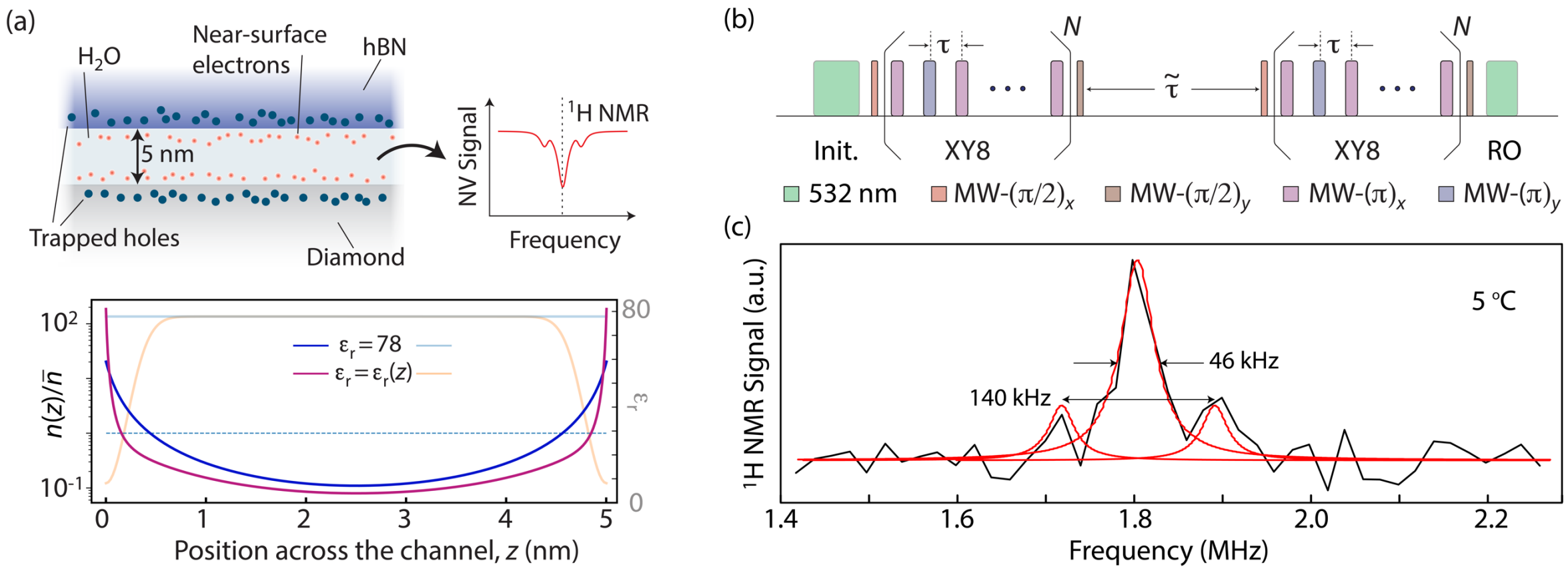

**Fig. 3: Modeling hyperfine signatures.** (a) Calculated equilibrium density profile of solvated excess electrons across a 5 nm water channel, showing charge accumulation near the channel walls for a uniform or variable relative dielectric constant $\varepsilon_r$ (faint traces). Protons located in different regions of the channel experience different hyperfine fields, leading to a structured proton NMR spectrum (upper schematic). (b) NV correlation spectroscopy protocol. The sequence effectively compares the proton-induced NV phase shifts upon a variable evolution time $\tilde{\tau}$. RO: Readout; Init.: Initialization. (c) Fourier transform of the signal acquired in (b) for $N = 5$ together with a schematic decomposition (red) into central and hyperfine-shifted contributions.

The appearance of a proton doublet of this magnitude is highly unusual in liquid-state water. Conventional chemical shifts arise from variations in electronic screening and typically produce frequency offsets in the few parts-per-million range, too small for $^1$H NMR observation at 35-50 mT. Likewise, direct proton-proton dipolar interactions are strongly averaged by molecular motion in liquids[26] even if, as in the present case, molecular diffusion is slower than normal. Further, crystalline ice—which, under special conditions, can yield a Pake-like pattern[27]—is difficult to reconcile with the observations, as the gradual loss of signal in Fig. 1f is most naturally interpreted as evaporation of liquid water, the splitting persists up to 14 $^{\circ}$C, and the spectrum would require the ice c-axis to be virtually colinear with the observed NVs (forming a ~54.7 deg. angle with the surface normal). These considerations point toward a different origin, one possibility being the coupling of protons to proximal electronic spins.

Hyperfine interactions have been invoked before[11], motivated in part by the fact that continuous optical excitation can generate mobile charge carriers within the device. In particular, illumination of the diamond sensor is known to produce photoexcited electrons that can be transferred to nearby environments[28-31], and similar processes likely apply to hBN as well. If the doublet reported here indeed originates from a hyperfine interaction, however, the observation immediately imposes several constraints on the underlying microscopic water state. First, the absence of substantial inhomogeneous broadening—key to our ability to resolve the doublet—implies that the effective hyperfine field experienced by all protons in the channel must be relatively uniform. Second, rendering the NMR splitting observable requires that fluctuations of the underlying electronic-spin environment occur on timescales longer than the inverse (effective) coupling constant, here corresponding to tens of microseconds or more. Finally, a splitting of ~30 kHz implies a minimum density of electronic spins sufficient to generate an average hyperfine field of comparable magnitude throughout the confined liquid. We elaborate on these conditions below.

We start with the requirement of a spatially homogeneous hyperfine field as it is particularly restrictive. In a nanoconfined geometry, one would normally expect charged species to accumulate near the oppositely charged interfaces, producing strong spatial variations in both charge density and hyperfine coupling. To determine whether a more extended distribution is possible, we model the mobile paramagnetic charge complexes as a one-dimensional density $n(\mathbf{r}) = n(z)$ across the channel height ($0 < z < a$). In the presence of a positive charge density distribution $\rho_+(\mathbf{r})$ on the walls, and using $e$ to denote the electronic unit charge, we define $\rho(\mathbf{r}) = \rho_+(\mathbf{r}) - e\,n(\mathbf{r})$, and obtain the equilibrium charge profile by minimizing the Helmholtz free energy

$$\mathcal{F}(n) = \mathcal{F}_{\mathrm{E}} + \mathcal{F}_{\mathrm{C}} + \mathcal{F}_{\mathrm{Ch}}\,. \quad (1)$$

In the above expression, $\mathcal{F}_{\mathrm{E}} = k_{\mathrm{B}}T\int d^3r\, n(\mathbf{r})(\ln(n(\mathbf{r})/n^*) - 1)$ describes the configurational entropy of the solvated electrons, and $\mathcal{F}_{\mathrm{C}} = \frac{1}{2}\int d^3r' \int d^3r \frac{\rho(\mathbf{r})\rho(\mathbf{r}')}{4\pi\epsilon_r\epsilon_0|\mathbf{r}-\mathbf{r}'|}$ accounts for their attraction to the fixed positive charges within the channel walls together with their mutual electrostatic repulsion. Lastly, $\mathcal{F}_{\mathrm{Ch}} = \int d^3r\, n(\mathbf{r}) U_{\mathrm{Ch}}(\mathbf{r})$ incorporates non-electrostatic hydration and solvation effects, with $U_{\mathrm{Ch}}(\mathbf{r})$ quantifying the energy cost of placing a charge complex at position $\mathbf{r}$.

We first neglect $\mathcal{F}_{\mathrm{Ch}}$ and ask whether electrostatics and entropy alone can support a charge distribution consistent with the observed NMR response. The corresponding equilibrium charge distribution is obtained by minimizing the free-energy functional subject to charge conservation (SM, Section III). Independent of whether the positive charges are assumed to be uniformly distributed or localized into patches within the channel walls, the equilibrium solution consistently predicts accumulation of the solvated charge complexes near the water–diamond interfaces (Fig. 3a); we find the same trend regardless the setback distance assumed for $\rho_{+}$. Further, introducing a position-dependent relative dielectric response $\epsilon_{\rm r}(z)$ with reduced screening near the channel walls[16,32,33], tends to enhance interfacial localization.

The above distribution gives rise to a heterogeneous environment, i.e., protons located near the channel walls experience appreciable hyperfine fields, whereas those in the channel interior remain only weakly perturbed. The simplest expectation is therefore a structured proton NMR spectrum consisting of a central resonance together with hyperfine-shifted satellite contributions, as schematically illustrated in Fig. 3a. To probe the predicted spectral heterogeneity, we use NV correlation spectroscopy[18,34], which separates proton populations through their distinct phase evolution during the variable interval $\tilde{\tau}$ (Fig. 3b). The Fourier-transformed signal reveals weak shoulders approximately 70 kHz on either side of the central proton resonance, consistent with a minority population experiencing substantially larger hyperfine fields than the remainder of the confined liquid (Fig. 3c and SM, Section II).

In light of these observations, a natural question is whether the transient doublet in Fig. 2 could simply represent an extension of this same picture in which a new class of hydration complexes—now featuring a weakened $A_0{\sim}30$ kHz hyperfine coupling—become dominant throughout the channel. Within the framework of Eq. (1), the transient state may emerge when the hydration contribution $\mathcal{F}_{\rm Ch}$ of a spatially extended state lowers the free energy sufficiently to compete with the electrostatically favored interfacial state. Near-degeneracy between these contending minima may account for the observed intermittency, with modest perturbations—including temperature fluctuations or stochastic changes in the distribution of trapped charges on the channel walls—shifting the balance. Notice that the emergence of a doublet is still conceivable in a heterogeneous charge distribution if the number of protons in contact with the wall-clustered electrons is too low to yield an observable NMR signal, or if strong hyperfine shifts render their resonances effectively undetectable.

The extracted value for $A_0$ is significant in that it sheds some light on the implied nature of the hydration complexes at play. Indeed, taking characteristic local electron-proton hyperfine couplings of ~2 MHz[35,36], the observed effective coupling suggests an average over a nanometer-scale hydration volume containing tens of water molecules. This spatial averaging must be faster than the nuclear Larmor frequency so as to avoid a broad distribution of proton shifts. At the same time, it should involve a single electronic spin projection since the converse implies a fluctuating sign of the shift. Thus, the presence of an NMR doublet requires a spin-bearing charge-hydration configuration that is spatially extended over a ~1-nm-size molecular ensemble and remains correlated for at least $A_0^{-1}$, i.e., tens of microseconds.

Importantly, the near-complete conversion of the proton resonance into a doublet leaves little residual spectral weight at the original resonance frequency, implying that most protons contributing to the NV signal experience a comparable hyperfine interaction. As the separation between neighboring charge-hydration complexes becomes comparable to their size, they may cease to behave independently and instead form an extended, correlated network within the confined liquid. In this sense, the hypothesized state may bear some resemblance to Coulomb-dominated electronic phases[37], although here the ordering would be intrinsically coupled to the structure and dynamics of the liquid itself. Further, one can plausibly envision a family of related charge-hydration complexes differing in size, local hydration structure, and charge density. Since the effective hyperfine coupling depends on how the electronic spin density is distributed across the surrounding proton network, modest variations in these parameters would naturally produce the range of splittings observed experimentally (≈20–40 kHz). The associated reorganization of the surrounding water molecules may further reduce the effective Coulomb interaction with the channel walls through enhanced dielectric screening, providing a microscopic mechanism by which the hydration contribution can compete with electrostatic localization.

Intriguingly, recent theoretical work has suggested that a solvated electron can reorganize the surrounding liquid over a nanometer-scale region encompassing several tens of water molecules, inducing both orientational and translational correlations within the hydrogen-bond network[38]. Within this perspective, the confined liquid studied here may represent the many-body extension of this picture, where neighboring hydration complexes cease to behave independently and instead may collectively organize through their Coulomb interactions. The observed intermittency would then reflect the reversible formation and decay of these extended, charge-stabilized correlated domains.

Lastly, we note that while a nearly uniform distribution of solvated charge may be necessary to stabilize the long-lived correlated state hypothesized here, it need not be the direct origin of the observed spectral splitting. An alternative possibility is that the excess electrons serve primarily to stabilize the correlated state while remaining localized within individual hydration complexes[39-42]. Their large associated hyperfine couplings would then become unobservable, and the measured doublet would instead arise from residual proton-proton dipolar interactions. Such a scenario, however, would require an unusual dynamical regime in which intramolecular dipolar couplings retain a preferred orientation while intermolecular couplings are efficiently averaged by correlated molecular motion[43]. Distinguishing between these microscopic mechanisms will require further theoretical and experimental investigation involving the use of channel structures of varying height, controlled tuning of interfacial charge density, and ab initio-informed molecular dynamics of solvated charge complexes under confinement.

We thank Dieter Suter and Friedemann Reinhard for helpful discussions. R.K., K.X., N.J. and D.P., C.A.M. acknowledge support by the National Science Foundation through grant NSF-2506082. R.B. and A.B. acknowledge funding from the European Union's H2020 Framework Programme/ERC Starting Grant 852674 AngstroCAP, Royal Society University Research Fellowship URF\R\231008, Philip Leverhulme Prize PLP-2021-262, EPSRC new horizons grant EP/X019225/1, and EPSRC strategic equipment grant EP/W006502/1. All authors acknowledge access to the facilities and research infrastructure of the NSF CREST IDEALS, grant number NSF-2112550.

# Supplementary Material for

# "Observation of long-lived spin order in nanoconfined water"

Rohma Khan[1,2], Kang Xu[1], Nathaniel Jeffries[1], Ankit Bhardwaj[3], Boya Radha[3,4], Daniela Pagliero[1,*], and Carlos A. Meriles[1,2,*]

[1]*Department of Physics, CUNY–The City College of New York, New York, NY 10031, USA.*
[2]*CUNY-Graduate Center, New York, NY 10016, USA.*
[3]*Department of Physics and Astronomy, The University of Manchester, Manchester, UK.*
[4]*Photon Science Institute and National Graphene Institute, University of Manchester, UK.*

[*]Corresponding authors: dpagliero@ccny.cuny.edu, cmeriles@ccny.cuny.edu

## I. Materials and experimental methods

### I.1 *Device architecture and temperature-control platform*

The nanochannel devices employed in this work are based on the refillable hBN nanochannel platform on diamond described previously in Ref. [1]. Briefly, 275-nm-wide hBN spacers in the form of long strips were used to define 185-nm-wide channels of nominal height ~5.6 nm between a diamond sensing substrate and a hBN cover layer; the latter is sufficiently thick (70 nm) to avoid sagging or bulging. The resulting structure was integrated into a microfluidic chamber that enabled repeated filling and refilling cycles throughout the course of the experiments. To this end, we circulated a saturated water vapor through the sample chamber; water accesses the channels from windows patterned via dry etching near the ends of the spacer strips.

To investigate the temperature dependence of the confined water, the pristine face of the diamond crystal was mounted on a thermoelectric (Peltier) stage providing active temperature control between room temperature and near-freezing conditions. A thermistor placed in close proximity to the nanochannel chip continuously monitored the stage temperature during operation; we also used the known thermal dependence of the NV zero-field splitting to determine the diamond temperature in-situ (see below). To prevent condensation of ambient humidity, the microscope head and sample assembly were enclosed, and a continuous flow of dry nitrogen (or dry air) was directed onto the sample chamber through a dedicated nozzle. A schematic of the experimental assembly is shown in Fig. S1a; images of the microfluidic device and the entire assembly including the mw antenna — a double, 0.5-mm-diameter loop of thin coated copper wire glued to one of the windows — are presented in Fig. S1b.

### I.2 *NV magnetometry and thermometry*

The nanochannel structure was fabricated on a [100]-oriented electronic-grade diamond (Element Six) with dimensions of 2×2×0.5 mm$^3$. To reduce the $^{13}$C spin background, the substrate was overgrown with a 1-µm-thick epilayer enriched to 99.99% $^{12}$C, after which the surface was polished to an average roughness below 1 nm. A shallow ensemble of NV centers was then generated by implanting $^{15}$N ions (3 keV, $1\times10^{12}$ cm$^{-2}$) and annealing the crystal at 850 ºC. The resulting sensing layer was located approximately 8 nm below the diamond surface and contained an estimated density of ~$10^2$ NV centers per µm$^2$. Prior to fabrication of the nanofluidic device, the surface was oxygen-terminated using a standard acid-cleaning procedure. Upon incorporating the nanochannels, the device was cleaned only by acetone immersion to avoid compromising the nanostructures.

Measurements were carried out with a home-built confocal microscope equipped with 532 nm optical excitation and microwave control of the NV electronic spin transitions. We used a permanent magnet to produce a field in the range 35-50 mT along a selected NV axis. The magnetic field was chosen so as to operate near the excited-state level anticrossing (~51.2 mT), enabling efficient polarization of the host $^{15}$N nuclear spin. Under these conditions, ODMR linewidths of approximately 2 MHz were routinely obtained, substantially narrower than typically observed in dense NV ensembles. XY8-*N* pulse sequences were used to render the NV spin sensitive to oscillating magnetic fields generated by nearby proton spins. Resonant detection occurs when the interpulse spacing satisfies the condition that twice the pulse spacing matches the proton Larmor period. NMR spectra were obtained by sweeping the pulse spacing and recording the corresponding NV spin contrast.

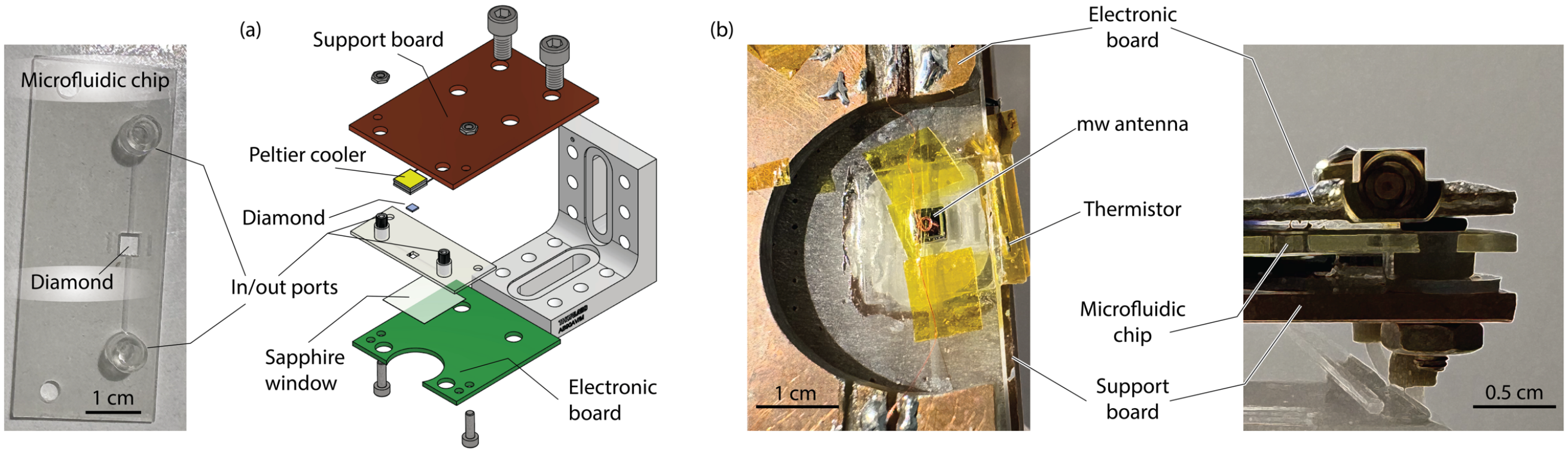

**Fig. S1: Temperature-controlled platform for nanoscale NV–NMR of confined water.** (a) Photograph of the microfluidic chip used in this work (left) and exploded view of the temperature-controlled sample assembly (right). The chip consists of a diamond substrate supporting the nanochannel device, with fluid access provided through inlet and outlet ports. Temperature control is achieved using a Peltier cooler mounted beneath the sample, while optical access is provided through a sapphire window. (b) Top and side views of the assembled platform installed on the microscope stage. The microfluidic chip is mounted on a support board containing a thermistor for independent temperature monitoring. Microwave control of the NV centers is provided by striplines patterned on the electronic board positioned above the sample (inner side of the board, not visible in the photo). To prevent condensation of ambient humidity, we continuously blow dry air or nitrogen through an enclosure surrounding the microscope head (not shown). The assembly enables simultaneous optical interrogation, microwave control, fluid delivery, and active temperature regulation during NV-NMR measurements.

Temperature measurements were obtained via a thermistor or directly from the NV center zero-field splitting. ODMR spectra were acquired simultaneously on the $|0\rangle \leftrightarrow |+1\rangle$ and $|0\rangle \leftrightarrow |-1\rangle$ transitions, and the average transition frequency $\Delta = (\nu_+ + \nu_-)/2$ was used to remove the first-order Zeeman contribution. The resulting value of $\Delta$ was converted into temperature using the known temperature dependence of the NV zero-field splitting[2].

Prior to measurement, the hBN nanochannels were filled with water by exposing the device to a saturated water-vapor environment, allowing capillary condensation to spontaneously fill the channels. For the experiments presented in Figs. 2b and 2c of the main text, however, the microfluidic chamber was instead kept continuously flooded with deionized water to suppress evaporation during the extended measurements (see below). A slow flow (6 μL min$^{-1}$) was maintained throughout the experiment to prevent bubble formation within the chamber. At this flow rate, the total volume of water in the microfluidic chamber was replaced only approximately every 1.5 days, ensuring that the flow itself minimally perturbed the confined liquid.

## II. Additional spectroscopic characterization of the confined water

### II.1 *Liquid-state persistence and diffusion suppression*

The observation of an $^1$H NMR resonance from the confined water via NV centers relies on the strongly suppressed molecular diffusion previously reported for this system. As discussed in Ref. [1], molecular dynamics simulations predict that confinement within a $\sim$5 nm channel should produce nearly bulk-like transport and thus lead to negligible NV-NMR signal because rapid molecular diffusion broadens the proton resonance beyond detection as spins traverse the shallow NV sensing volume. Since geometric confinement alone cannot account for the observed slowdown, the reduced diffusivity was attributed to the presence of photoinduced charge within the nanochannels, which modifies the local electrostatic environment and hinders molecular transport. The present measurements (Figs. 1 and 2 in the main text) demonstrate that this diffusion-suppressed liquid state persists throughout the investigated temperature range.

The proton resonance linewidth exhibits little systematic variation with temperature (Fig. 1e in the main text). Following the analysis of Ref. [1], the effective translational diffusion coefficient can be estimated from the measured linewidth by equating the diffusion time across the characteristic sensing radius, $\tau_{\mathrm{D}} \approx L^2/(6D)$, to the inverse linewidth. Using $L \approx 8$ nm and $\Delta\nu \approx (\pi\tau_D)^{-1} \approx 40$ kHz yields an effective diffusivity of order $10^{-12}$ m$^2$ s$^{-1}$, confirming that the strongly suppressed molecular dynamics remain essentially unchanged over the explored temperature range.

The observations discussed in the main text rely on the assumption that the confined water remains in a liquid state throughout the experiment. Figure S2 provides support for this interpretation by tracking the proton NMR signal over several weeks as water gradually evaporates from the refillable nanochannel device. We find that the signal amplitude decreases continuously with time owing to the reduction in the amount of water within the channels. Importantly, however, the spectral linewidth remains approximately unchanged throughout the measurement, despite the substantial loss of signal and the progressive reduction in temperature.

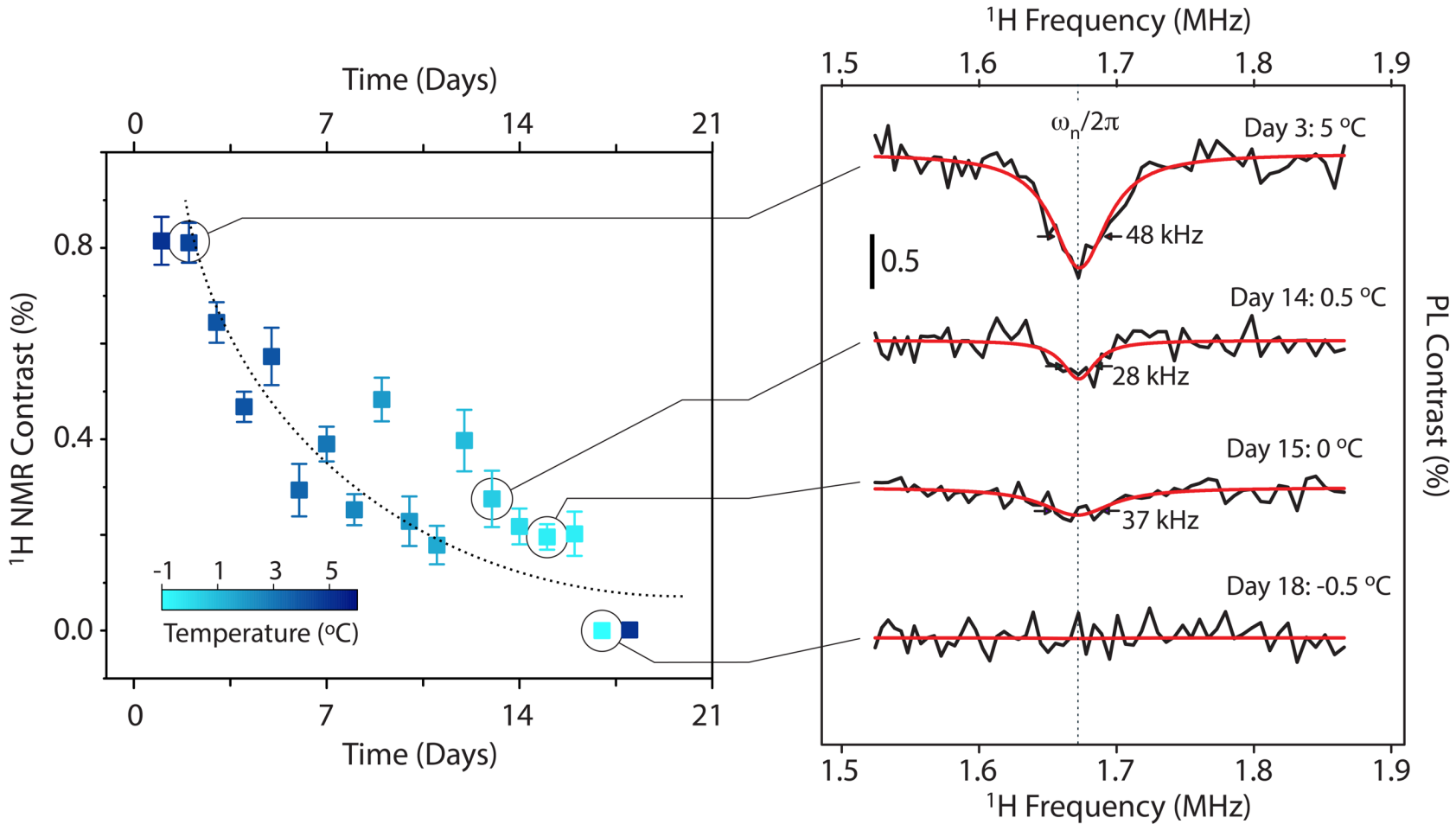

**Fig. S2: Long-term evolution of the confined-water $^1$H NMR signal.** Left (reproduced from Fig. 1f): $^1$H NMR contrast as a function of time following channel filling. The gradual decrease in signal reflects progressive evaporation of water. Right: Representative spectra at selected times. Although the signal amplitude decreases as water is lost, the linewidth remains approximately unchanged throughout the experiment. The proton resonance stays observable down to temperatures near and slightly below 0 ºC, consistent with the persistence of a liquid phase under confinement. Red curves are Lorentzian fits.

The simplest interpretation is that evaporation removes water from portions of the nanochannel network without significantly modifying the properties of the regions that remain filled (i.e., without leaving only strongly interfacial water behind). The near invariance of the spectral lineshape therefore suggests that the water contributing to the measured signal continues to occupy channels that remain fully liquid-filled, even as the total volume of confined water decreases.

### II.2 *Evidence for strongly hyperfine-coupled proton populations*

The proton spectra discussed in the main text are primarily obtained from measurements sensitive to the dominant proton population within the channels. Additional information can be obtained from correlation spectroscopy, which probes longer timescales and provides improved spectral resolution. Figure 3c in the main text shows a representative measurement acquired at 5 °C using the correlation protocol described in the inset. The resulting spectrum consists of a dominant central proton resonance together with two weaker satellite peaks, yielding an apparent triplet structure. A fit to the spectrum gives a central linewidth of approximately 46 kHz and a satellite splitting of approximately 140 kHz. Notably, the satellite resonances are substantially narrower than the central line, with linewidths of order 25 kHz. This narrowing suggests that the corresponding proton population experiences reduced motional averaging and therefore slower local dynamics than the majority of the confined liquid.

The observed triplet is reminiscent of features reported previously in NV-NMR measurements of nanoconfined water at room temperature, where proton splittings approaching 190 kHz were observed[1]. In both cases, the large spectral separations point to proton populations experiencing substantially stronger local hyperfine fields than those responsible for the dominant resonance. A natural interpretation is that these protons reside near the confining interfaces, where photoinduced charge complexes generated at or near the channel walls can produce stronger hyperfine interactions than those experienced by molecules occupying the channel interior.

It is important to distinguish the spectral features observed in Fig. 3 from the transient doublet in Fig. 2 (see also Figs. S3 and S4 below). The correlation measurements require acquisition times extending over several days in order to accumulate sufficient signal-to-noise ratio. Consequently, the measured spectrum represents a long-time average over many microscopic configurations. Because the spectral lineshape is not known a priori and may evolve during the measurement, conventional undersampling approaches commonly employed in correlation spectroscopy cannot be applied without risking spectral distortions. As a result, transient features that appear and disappear on timescales of hours, such as the ~20–40 kHz doublet reported in the main text, are expected to average out in the correlation measurement. In contrast, persistent proton populations experiencing strong local hyperfine fields remain visible, giving rise to the satellite peaks observed in Fig. 3c of the main text.

### II.3 *Intermittent appearance of the split proton resonance*

As discussed in the main text, the most unusual feature observed in the present study is the occasional appearance of a split proton resonance. Figure S3 presents additional examples of spectra exhibiting this behavior, including measurements performed at different temperatures. Although the precise splitting varies somewhat between observations, with values ranging from approximately 20 to 40 kHz, the qualitative spectral signature remains the same, i.e., a proton resonance that resolves into two comparably weighted components separated by several tens of kilohertz. The variation in the observed splitting may itself contain information about the microscopic state of the system. Within the interpretation discussed in the main text, different splitting values could reflect modest variations in the density, spatial distribution, or microscopic character of the underlying charge-hydration complexes.

At present, the conditions governing the appearance of the doublet remain incompletely understood. Measurements performed under nominally similar experimental conditions often yield conventional single-line spectra, while a subset reveal a resolved splitting. This intermittency suggests that the doublet is associated with a metastable configuration of the confined liquid, not a permanent characteristic of the nanochannel environment.

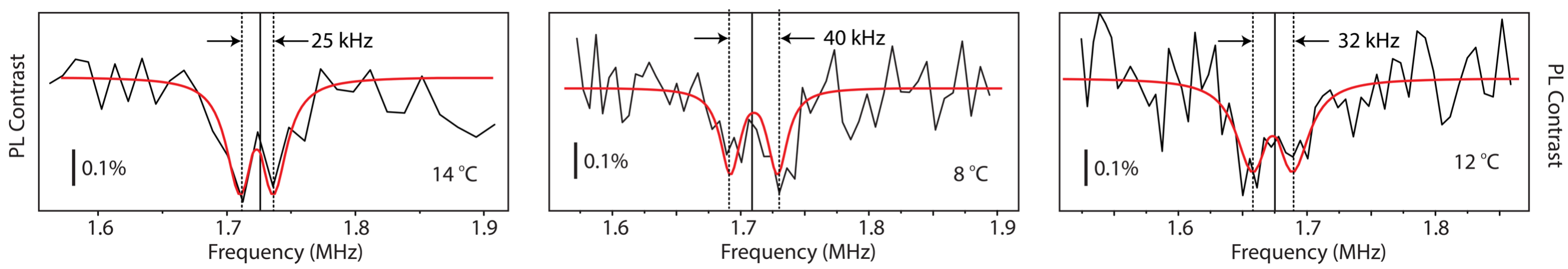

**Fig. S3: Additional $^1$H NMR spectra exhibiting a resolved doublet under different experimental conditions.** Although the majority of doublet observations occurred within a narrow temperature window near 4−5 °C, the examples shown here indicate that the phenomenon is not restricted exclusively to that range. The solid vertical line marks the expected proton Larmor frequency, while the dashed lines indicate the positions of the two doublet components. The red curves serve as guides to the eye.

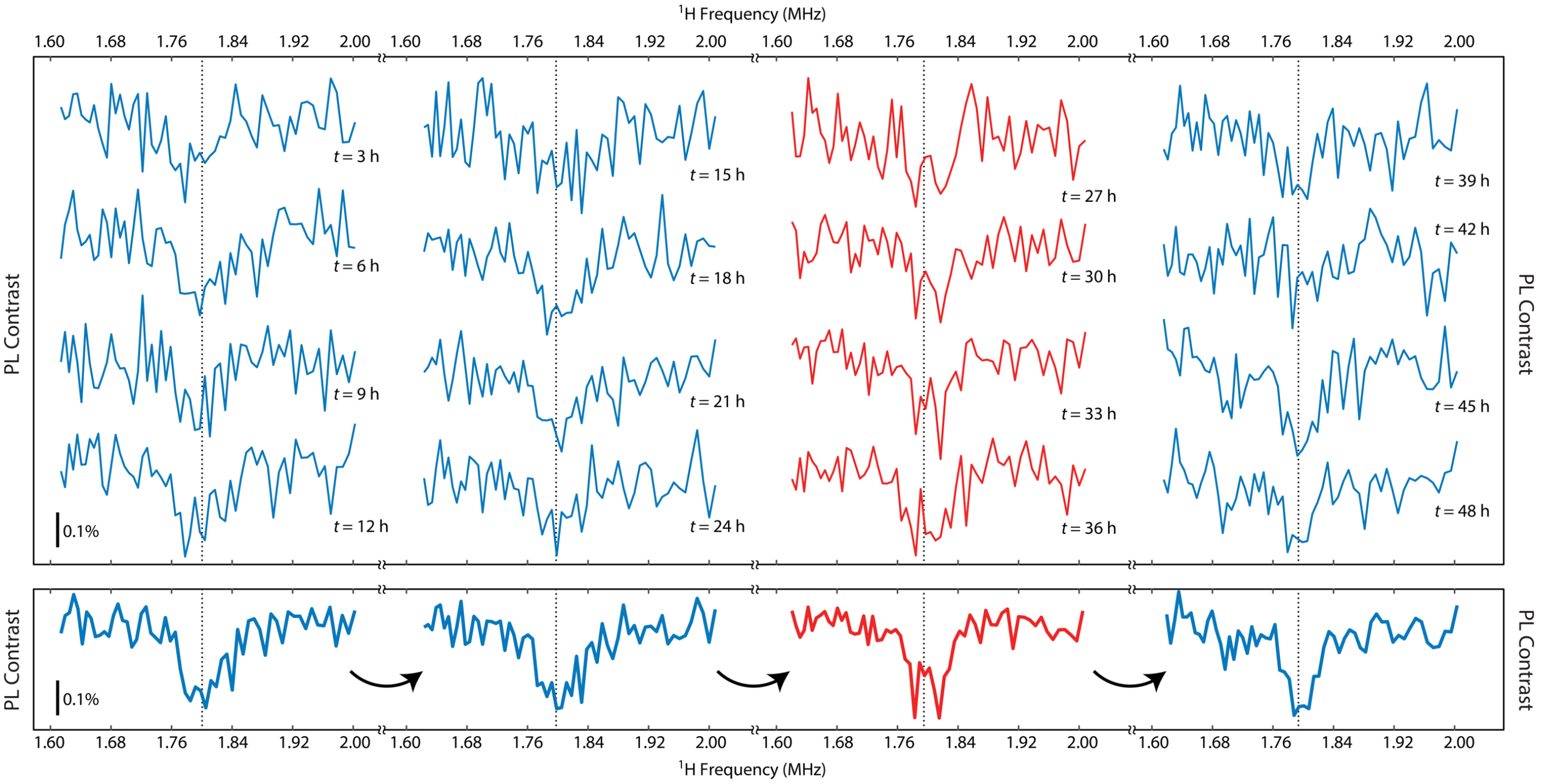

**Fig. S4: Temporal evolution of the split proton resonance.** Top: Consecutive $^1$H NMR spectra acquired over a 48 h interval while maintaining the sample near 5 ºC. Each spectrum is obtained from the shortest acquisition window compatible with the available signal-to-noise ratio, allowing the evolution of the lineshape to be monitored in time. Spectra exhibiting a resolved doublet are highlighted in red. Bottom (reproduced from Fig. 2b): Averages of the spectra within each time interval, showing the progression from an initially single resonance (0–24 h), to a transient doublet state (27–36 h), and finally a return to a predominantly single resonance (39–48 h). The vertical dashed lines mark the frequencies of the split peaks observed during the intermediate interval. The gradual emergence and disappearance of the doublet demonstrate that the intermittency reported in the main text originates from a long-lived but transient configuration of the confined liquid.

One possible contributing factor is the repeated optical excitation used throughout the experiment. Each measurement cycle involves illumination with green light, generating carriers in the diamond and periodically modifying the electrostatic environment experienced by the confined water. The resulting redistribution of charge, together with the formation of space-charge fields and local field gradients arising from the spatially inhomogeneous illumination profile, may gradually alter the configuration of charge complexes within and around the nanochannels. Such changes could in turn modify the stability of the confined liquid and potentially trigger transitions between the conventional single-line state and the split-spectrum configuration. Determining the origin of this variability will require future experiments capable of controlling and monitoring the charge environment of the nanochannels more directly.

### II.4 *Temporal evolution of the doublet state*

To better understand the origin of the observed intermittency, we monitored the proton spectrum continuously over extended periods while maintaining the sample near 5 ºC; to avoid evaporation, the microfluidic chamber was kept continuously flooded with deionized water, with a slow flow (6 μL min$^{-1}$) maintained to prevent bubble formation (see above, Section I.2). The upper plot in Fig. S4 shows the dataset used to derive the order parameter Φ presented in the main text. The full acquisition was divided into a sequence of shorter time windows, and each resulting spectrum was fitted using both a single-resonance model and a doublet model. We then define an order parameter $\Phi = \chi_s^2/\chi_d^2$, where $\chi_s^2$ ($\chi_d^2$) denote the residual sums of squares obtained from the single (double) peak fit. Spectra highlighted in red correspond to time intervals for which the extracted value of Φ indicates preference for a doublet lineshape.

To assess the robustness of this metric, the analysis was repeated using several fitting protocols, including models with different combinations of free and constrained line-shape parameters (e.g., peak positions, linewidths, and amplitudes). The different fitting strategies yield slightly different absolute values of Φ, but all exhibit the same pronounced transient increase followed by a gradual return to baseline (Fig. 2c in the main text).

The data reveal that the split spectrum appears relatively abruptly on the timescale accessible to the measurement. No evidence of a doublet is visible in the spectrum acquired at $t = 24$ h, whereas a clearly resolved splitting is present only three hours later, at $t = 27$ h. The doublet remains evident in the spectra acquired at $t = 30$, 33, and 36 h, and may still be weakly present at $t = 39$ h before disappearing in the subsequent measurements. Averaging the spectra within successive time intervals emphasizes this behavior, revealing a well-defined period during which the split state dominates the observed lineshape (lower plot in Fig. S4). The timescale associated with this evolution is particularly noteworthy: Within the framework proposed in the main text, the data point to the transient formation of a singular, solvated-electron-stabilized state that eventually relaxes back to a more conventional configuration where electrons cluster near the walls.

The microscopic origin of the transition remains unclear. One possible contributing factor is the repeated optical excitation employed throughout the experiment. Each measurement cycle involves illumination with green light, generating carriers in the diamond and modifying the electrostatic environment experienced by the confined water. Over many hours, the resulting redistribution of charge, together with the formation of space-charge fields and local field gradients arising from the spatially inhomogeneous illumination profile, may gradually alter the configuration of charge complexes within and around the nanochannels. Within this picture, the appearance and disappearance of the doublet would reflect transitions between distinct charge-hydration configurations characterized by different degrees of spatial correlation and hyperfine homogeneity.

Although the present measurements do not reveal the microscopic mechanism responsible for the transition, they provide evidence that the split spectrum is associated with a slowly evolving state of the nanoconfined system. Thus, any successful microscopic description must account not only for the observed hyperfine splittings, but also for their transient persistence on timescales extending to several hours.

## III. Thermodynamic model and microscopic constraints imposed by the hyperfine doublet

### III.1 *Physical motivation*

The observation of a resolved proton doublet imposes constraints that go beyond the presence of photoinduced paramagnetic charge carriers in the fluid. If the splitting does originate from hyperfine interactions — which, as discussed in the main text, appears to be the only plausible explanation — the effective hyperfine field must be sufficiently homogeneous over the proton ensemble detected by the NV centers. In addition, the microscopic configuration responsible for the effective hyperfine field must remain correlated for a time longer than the inverse splitting, as the opposite implies motional averaging of the two branches into a single resonance.

This section expands on the minimal model discussed in the main text for the spatial distribution of mobile paramagnetic charge complexes in the confined liquid, and derives the constraints imposed by the measured splitting. We emphasize that the goal is not to identify a unique microscopic charge species, but to determine what class of microscopic states is compatible with the observed NMR response.

### III.2 *Helmholtz free-energy functional*

We model the confined liquid as a slab of height $a$, with coordinate $z$ in $[0, a]$ measured from the diamond-water interface. Mobile negative paramagnetic charge complexes are described by a number density $n(\mathbf{r})$, while fixed positive charge in the confining walls is represented by a prescribed density $\rho_+(\mathbf{r})$. The total charge density is

$$\rho(\mathbf{r}) = \rho_+(\mathbf{r}) - en(\mathbf{r})\,, \tag{1}$$

where $e > 0$ is the elementary charge. At fixed total number of mobile complexes, $N = \int n(\mathbf{r})d^3r$, the equilibrium density profile is obtained by minimizing the Helmholtz free energy

$$\mathcal{F}(n) = \mathcal{F}_\mathrm{E} + \mathcal{F}_\mathrm{C} + \mathcal{F}_\mathrm{Ch}\,. \tag{2}$$

The entropic contribution at temperature $T$ is taken as the ideal configurational entropy of the mobile charge complexes, i.e.,

$$\mathcal{F}_\mathrm{E} = k_\mathrm{B}T\int dr^3\, n(\mathbf{r})\left(\ln\left(\frac{n(\mathbf{r})}{n^*}\right) - 1\right) \tag{3}$$

where $n^*$ serves as a reference for the chemical potential, and $k_\mathrm{B}$ denotes as usual, the Boltzmann constant. The electrostatic contribution is

$$\mathcal{F}_\mathrm{C} = \frac{1}{2}\int d^3r' \int d^3r\, \rho(\mathbf{r})\rho(\mathbf{r}')G(\mathbf{r},\mathbf{r}') \tag{4}$$

where $G(\mathbf{r},\mathbf{r}')$ denotes Green's function. For the minimal calculations in the main text we use a uniform effective dielectric constant, implying $G(\mathbf{r},\mathbf{r}') = \frac{1}{4\pi\varepsilon_\mathrm{r}\varepsilon_0|\mathbf{r}-\mathbf{r}'|}$. More generally, $G$ may be replaced by the solution of the Poisson

equation with a position-dependent dielectric response, $\varepsilon_{\rm r}(\mathbf{r})$.

The electrostatic potential generated by the total charge density is $\phi(\mathbf{r}) = \int d^3r' \, \rho(\mathbf{r}')G(\mathbf{r},\mathbf{r}')$. With this definition, the electrostatic free energy may be written equivalently as

$$\mathcal{F}_{\rm C} = \frac{1}{2}\int d^3r\rho(\mathbf{r})\phi(\mathbf{r}) \, . \tag{5}$$

Finally,

$$\mathcal{F}_{\rm Ch} = \int d^3r \, n(\mathbf{r})U_{\rm Ch}(\mathbf{r}) \tag{6}$$

accounts phenomenologically for non-electrostatic solvation, hydration, or interfacial binding effects. A positive $U_{\rm Ch}$ near the walls represents an interfacial hydration penalty for placing a charge complex close to a solid surface, while a negative $U_{\rm Ch}$ would describe specific adsorption.

The Euler-Lagrange equation follows from minimizing

$$\tilde{\mathcal{F}}(n) = \mathcal{F}(n) - \mu\int d^3r \; n(\mathbf{r}) \tag{7}$$

where $\mu$ enforces fixed total mobile charge. The resulting condition is $k_{\rm B}T\ln(n(\mathbf{r})/n^*) - e\phi(\mathbf{r}) + U_{\rm Ch}(\mathbf{r}) - \mu = 0$. Equivalently,

$$n(\mathbf{r}) = n_0 \exp\left(-\frac{1}{k_{\rm B}T}\big((-e)\phi(\mathbf{r}) + U_{\rm Ch}(\mathbf{r}) - \mu\big)\right) . \tag{8}$$

This expression captures the competition between electrostatic attraction toward positive wall charge, configurational entropy, mutual charge repulsion through the self-consistent potential, and any non-electrostatic solvation terms.

**III.3 *One-dimensional reduction and numerical implementation***

For the calculations shown in the main text, we reduce the problem to a laterally averaged density profile $n(z)$ across the channel height. Assuming translational invariance in the lateral directions, the density is approximated as $n(r) \to n(z)$. The electrostatic potential likewise becomes a function only of the coordinate normal to the interfaces,

$$\phi(z) = \int dz' K(z,z')[\rho_+(z') - en(z')] \, , \tag{9}$$

where $K(z,z') = \int dxdyG(r,r')$ is the laterally averaged interaction kernel. The coupled equations for $n(z)$ and $\phi(z)$ were solved self-consistently on a one-dimensional discretized grid, subject to the global neutrality constraint $\int_0^a n(z)dz = N/S$ where $S$ is the channel area. The resulting description is equivalent to a mean-field Poisson-Boltzmann treatment supplemented by the phenomenological term $U_{\rm Ch}(z)$, with correlations beyond the self-consistent electrostatic field neglected.

The positive wall charge is placed symmetrically in the two confining walls, with a setback distance $d$ measured from the liquid-solid interface. The parameter $d$ captures the fact that the compensating positive charge need not reside exactly at the water interface; it may be distributed within the diamond, hBN, or interfacial defect layers. The model further assumes identical boundary conditions at the two confining interfaces and therefore does not distinguish between diamond-water and hBN-water interactions. This approximation is intended to isolate the generic effects of confinement, electrostatics, and entropy, while minimizing material-specific parameters whose values are presently unknown.

In the minimal model used for Fig. 3a in the main text, we set $U_{\rm Ch} = 0$, impose global charge neutrality, and use either a constant or $z$-dependent $\epsilon_{\rm r}$. The compensating positive charge is represented by two uniform sheets located symmetrically outside the liquid. For infinite planar sheets, the precise setback distance — here set to 1 nm — does not alter the electrostatic solution inside the channel and therefore does not constitute an independent model parameter. Lastly, note that a short-range positive $U_{\rm Ch}(z)$ near the walls — omitted in our numerical modeling — favors delocalization by penalizing incompletely hydrated charge complexes at the interfaces.

For the calculations presented here, the average density was fixed at one mobile paramagnetic charge complex per approximately fifty water molecules ($\bar{n} \approx 0.67$ nm$^{-3}$), with the compensating positive surface charge chosen to satisfy global neutrality. We note, however, that the qualitative behavior remains unchanged over a broad range of mobile charge concentrations. The resulting density profiles should therefore be regarded as representative rather than quantitatively predictive. In particular, the minimal model neglects steric exclusion, dielectric saturation, finite-size effects, and non-electrostatic hydration interactions, all of which are expected to limit the degree of interfacial charge accumulation predicted by the classical mean-field description.

### III.4 *Microscopic scale implied by $A_0$, dynamical constraints, and coverage of the sensed liquid volume*

The observed hyperfine splitting $A_0$ provides an estimate of the spatial scale over which an electronic spin must be distributed or dynamically averaged. Local electron-proton hyperfine couplings for protons in the immediate environment of a localized charge center can be of order ~2 MHz[3,4]. If the effective coupling measured by NMR is $A_0 \approx 30$ kHz, the electronic spin must be averaged over roughly $N_{\mathrm{eff}} \sim A_{\mathrm{loc}}/A_0 \sim 40-80$ local proton-coupling units. Because the coupling may not be uniform over all protons in the hydration environment, this estimate should be interpreted only as an order-of-magnitude estimate of the size of the associated charge-hydration domain. Using the molecular density of liquid water, $n_{\mathrm{H_2O}} \approx 33$ nm$^{-3}$, this corresponds to an effective volume $V_{\mathrm{eff}} = N_{\mathrm{eff}}/n_{\mathrm{H_2O}}$ containing tens of water molecules. Equivalently, the relevant radius $R_{\mathrm{eff}}$ is on the order of 1 nm, depending on the assumed microscopic coupling and weighting of nearby protons.

The doublet imposes two distinct dynamical constraints. First, the electronic spin density must be spatially averaged over a local proton ensemble. This can occur either because the electronic wavefunction is itself spatially extended over a hydration domain or because a spin-bearing charge complex samples several molecular configurations rapidly compared with the NMR timescale. Second, this spatial averaging must preserve the electronic spin projection as the alternative scenario — where protons sample many uncorrelated electronic spins during the NMR interrogation time — implies a fluctuating sign in the hyperfine shift and thus a collapse of the two branches of the doublet into a single averaged or broadened resonance. The spin-bearing entity must therefore be a charge-hydration configuration that is spatially extended over a nanometer-scale molecular ensemble while maintaining a correlated spin projection not shorter than $\tau_{\mathrm{corr}} \sim 1/A_0$; for $A_0 = 30$ kHz, $\tau_{\mathrm{corr}} \gtrsim 30$ µs. This lifetime is much longer than the characteristic timescales for molecular reorientation or hydrogen-bond rearrangement in bulk water, pointing instead to a long-lived charge-hydration configuration. Such behavior departs markedly from the conventional picture of rapidly diffusing solvated electrons or an ensemble of independent paramagnetic defects.

The presence of a doublet also constrains the spatial coverage of the sensed liquid volume. Protons that do not experience a significant hyperfine interaction would contribute an unsplit resonance at the center of the spectrum. The near-complete conversion of the $^1$H resonance into a doublet therefore indicates that nearly all protons contributing to the NV signal experience comparable effective hyperfine fields. This observation is difficult to reconcile with isolated paramagnetic complexes embedded in otherwise ordinary water. Instead, it suggests that the charge-hydration domains collectively occupy most of the sensed liquid volume, implying that neighboring domains are separated by distances comparable to their own spatial extent, a regime that hints at correlated electrostatic and hydration environments.

### III.5 *Temperature dependence*

Although split spectra are occasionally observed at other temperatures (Fig. S4), the apparent clustering of observations near 4–5 °C is particularly intriguing. One possible explanation is that this temperature range coincides with a regime where water exhibits anomalous structural and thermodynamic behavior associated with the competition between hydrogen-bond ordering and thermal disorder. Within the framework hypothesized here, the charge-hydration configurations responsible for the observed hyperfine fields are stabilized not only by electrostatic interactions but also by the surrounding hydrogen-bond network. It is therefore conceivable that the lifetime and spatial extent of these configurations become enhanced in this temperature range, allowing the emergence of the resolved doublet.

Other interpretations remain possible. For example, changes in the dielectric response of confined water could modify the balance between electrostatic attraction and screening, thereby affecting the formation and stability of extended charge distributions. Likewise, temperature-dependent variations in carrier trapping, detrapping, or transport at the diamond-water and hBN-water interfaces could alter the density of charge complexes available within the channel. Finally, the observed temperature window may reflect a kinetic effect rather than a thermodynamic one, corresponding to a regime in which the formation and decay timescales of the proposed charge-hydration configurations become comparable to the experimental observation time. Distinguishing among these possibilities will require future measurements with improved control over interfacial charge densities.

### III.6 *Resulting physical picture and limitations*

The magnitude of the hyperfine splitting, the inferred correlation time, and the near-complete conversion of the resonance into a doublet all point toward the transient formation of an extended spin-bearing charge-hydration state within the confined liquid. In this picture, photoinduced electronic charge does not remain localized near the interfaces but instead forms correlated charge-hydration configurations whose electrostatic and hydration environments overlap throughout much of the 5.6 nm channel. The resulting proton ensemble experiences a nearly homogeneous hyperfine interaction, giving rise to the observed doublet.

The inferred microsecond lifetime of the charge-hydration configurations may reflect their collective nature: rather than tracking the motion of individual water molecules, the relevant timescale would correspond to the reorganization of an extended, correlated charge-hydration network, analogous to how polarons or Coulomb-glass states evolve through cooperative rearrangements of their surrounding medium.

This interpretation remains phenomenological. The present model does not identify the microscopic charge carrier, determine its electronic wavefunction, or explicitly simulate hyperfine tensors in the confined liquid. The free-energy calculation models charge complexes as classical mobile species, while the hyperfine spectrum assumes a proportionality between local charge-complex density and effective proton splitting. These approximations establish consistency between the observed doublet and a spatially extended paramagnetic charge distribution, but they do not uniquely determine the microscopic structure.

A more complete theory would require multiscale simulations combining ab initio electronic-structure calculations of solvated or interfacial charge complexes with molecular dynamics of nanoconfined water under realistic electrostatic boundary conditions. Such calculations could establish whether spin-bearing charge-hydration configurations of nanometer-scale extent and microsecond-scale stability are physically plausible, and whether their predicted hyperfine fields match the observed splitting.